\documentclass[aps,prd,twocolumn,superscriptaddress,nofootinbib]{revtex4-1} %
\usepackage[utf8]{inputenc}
\usepackage[english]{babel}
\usepackage{amsmath}
\usepackage{amsfonts}
\usepackage{amssymb}
\usepackage{listings}
\usepackage{lipsum}
\usepackage{multirow}
\usepackage{datetime}
\usepackage{graphicx}
\usepackage{mathtools}
\usepackage{mathrsfs}
\usepackage{dcolumn}
\usepackage{multirow}
\usepackage{orcidlink}
\usepackage{color}   %May be necessary if you want to color links
\usepackage{hyperref}
\hypersetup{
    colorlinks=true, 
    pdfborder = {0 0 0.5 [3 3]},
    anchorcolor=black,
    citecolor=blue,
    linktoc=all,    
    linktocpage=true,
    linkcolor=red,
	urlcolor=blue
}

\newcommand{\red}[1]{{\color{black} {#1}}}

\begin{document}

\title{Nonlinear evolution of the ergoregion instability: \\ Turbulence, bursts of radiation, and black hole formation}
\author{Nils Siemonsen\orcidlink{0000-0001-5664-3521}}
\email[]{nils.siemonsen@princeton.edu}
%\homepage[]{Your web page}
%\thanks{}
%\altaffiliation{}
\affiliation{Princeton Gravity Initiative, Princeton University, Princeton, New Jersey 08544, USA}
\affiliation{Department of Physics, Princeton University, Princeton, New Jersey 08544, USA}
\author{William E.\ East\orcidlink{0000-0002-9017-6215}}
\email{weast@perimeterinstitute.ca}
\affiliation{Perimeter Institute for Theoretical Physics, Waterloo, Ontario N2L 2Y5, Canada}

\date{\today}

\begin{abstract} 
Spacetimes with an ergoregion that is not connected to a horizon are linearly unstable. While the linear regime has been studied in a number of settings, little is known about the nonlinear evolution of this ergoregion instability. Here, we investigate this by numerically evolving the unstable growth of a massless vector field in a rapidly spinning boson star in full general relativity. We find that the backreaction of the instability causes the star to become more gravitationally bound, accelerating the growth, and eventually leading to black hole formation. During the nonlinear growth phase, small scale features develop in the unstable mode and emitted radiation as nonlinear gravitational interactions mediate a direct turbulent cascade.
The gravitational wave signal exhibits bursts, akin to so-called gravitational wave echoes, with increasing amplitude towards black hole formation. 
\end{abstract}

\maketitle

\section{Introduction}
In general relativity, the rotation of an object affects its gravitational field,
giving rise to frame-dragging effects. As an extreme manifestation of this,
sufficiently compact and rotating objects have ergoregions: regions where the
asymptotically timelike Killing field becomes spacelike and no timelike
worldline remains stationary with respect to far away observers; test
particles or test fields in the ergoregion can carry negative Killing energy. 
As first pointed out by Penrose~\cite{Penrose:1971uk},
this property allows rotational energy to be liberated from a black hole,
including through the superradiant scattering of waves~\cite{zeldovic, Press:1972zz, Bardeen:1973gs}
or through the induced rotation of magnetic field lines in a black hole jet~\cite{1977MNRAS.179..433B}.

In a horizonless compact object, however, the existence
of an ergoregion not only allows the rotational energy
of the spacetime to be tapped into, but automatically gives rise to a linear instability.
As shown by Friedman~\cite{Friedman:1978} (see also Ref.~\cite{VILENKIN1978301}), without a horizon to fall through, 
a massless field fluctuation with negative Killing energy is trapped in the ergoregion.
By oscillating and emitting radiation carrying away positive energy, it can only 
grow a larger magnitude of negative energy, giving rise to a runway process---the ergoregion instability.

The instability growth rates have been estimated for
scalar~\cite{Comins:1978,Schutz:1978,Yoshida:1996} and gravitational wave (or
$w$-mode) perturbations of ultracompact neutron stars~\cite{Kokkotas:1992,
Kokkotas:2002sf} \red{(see also \cite{Tsokaros:2019mlz,Tsokaros:2020qju})}, as well as gravastars, boson stars (BSs), and
hypercompact Kerr-like objects \cite{Cardoso:2007az, Chirenti:2008pf,
Cardoso:2008kj,Maggio:2017ivp, Maggio:2018ivz,Siemonsen:2025wib}. \red{The growth timescales are typically many orders of magnitude longer than light crossing times of these objects. Hence, despite their linear instability, such objects could be formed dynamically, e.g. as remnants of compact binary mergers.}
Beyond this, the instability
was studied in the context of supergravity \cite{Cardoso:2005gj}, superspinars
\cite{Pani:2010jz}, braneworld scenarios \cite{Dey:2020pth}, in analogous
hydrodynamical systems \cite{Oliveira:2014oja, Hod:2014hda}, and in relation to
the superradiance instability \cite{Vicente:2018mxl}.  The presence of the
ergoregion instability was mathematically proved for the scalar wave equation
in Ref.~\cite{Moschidis:2016zjy}.

Despite the considerable amount of study in settings spanning high-energy physics to
gravitational wave astronomy, little is known about the nonlinear development
and ultimate endstate of the ergoregion instability. Recently, weakly nonlinear backreaction
was studied in Ref.~\cite{Siemonsen:2025wib}, finding evidence for a significant enhancement of the
unstable growth. Moreover, in Ref.~\cite{Siemonsen:2025fne} the evolution of an ergoregion unstable scalar field was studied without gravitational backreaction, and nonlinear scalar interactions were shown to result in a weakly turbulent direct cascade to small scales, leading to saturation.
If the generic backreaction of the instability was sufficiently
severe \red{(as opposed to saturating at a low level)}, it could be used to rule out the persistence of a whole class of horizonless objects
with properties similar to black holes, i.e. ``black hole mimickers" \cite{Cardoso:2019rvt,Carballo-Rubio:2025fnc,Bambi:2025wjx}.
As well, a nonlinear understanding of the ergoregion instability is essential to determining
whether there would be any observational signatures associated with this process
which could be used to probe such objects.
For example, in Refs.~\cite{Barausse:2018vdb,Fan:2017cfw}, it was argued that if the ergoregion
instability caused a population of ultracompact objects 
to spin down through gravitational wave emission, it could give rise
to a detectable gravitational wave background.

In this work, we seek to address the open question of what the backreaction of the
ergoregion instability is on the spacetime of a rotating compact object. Taking
a spinning, ultracompact BS as our model system, we use numerical
evolutions of the governing equations to follow the nonlinear development of a
unstable massless vector field within full general relativity. \red{To that end, we fix the dynamical spacetime to be axisymmetric, and enforce harmonic azimuthal dependencies in the matter fields. } We find that as 
the unstable field grows through the ergoregion instability, (i) nonlinear effects enhance 
its instability rate, (ii) the instability induces large oscillations in the star 
sourcing bursts of radiation with increasing amplitude, (iii) the presence of a 
gravitationally \red{mediated} turbulent cascade leaving imprints on the radiation, and (iv) that 
these nonlinear dynamics ultimately result in the collapse of the star to a rapidly spinning black hole.

\section{Methodology}

\begin{figure}[t]
\includegraphics[width=1\linewidth]{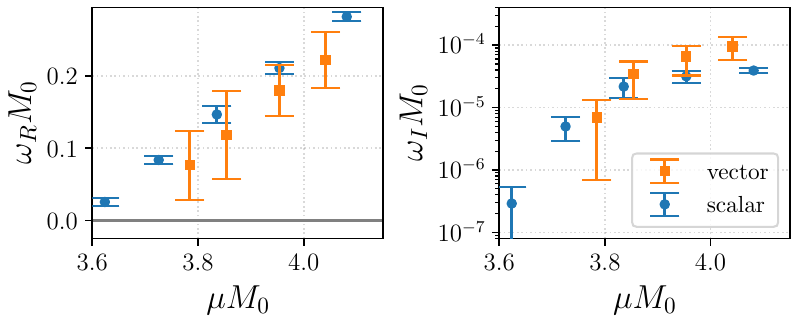}
\caption{Linear frequencies $\omega_R$ and growth rates $\omega_I$ of the most unstable $m=1$ massless vector test field configurations along the family of BS solutions with $\sigma=0.2$ and $\tilde{m}=3$. Error bars indicate the uncertainty of our numerical methods. The size of the ergoregion increases with $\mu M_0$. These are compared to the corresponding quantities for a massless scalar test field on this family of spacetimes obtained in Ref.~\cite{Siemonsen:2025wib}.}
\label{fig:vector_freq}
\end{figure}

We consider the theory with Lagrangian density
\begin{align}
\begin{aligned}
\mathcal{L}=& \ \frac{R}{16\pi}-\frac{|F|^2}{4} -|\partial\Phi|^2-\mu^2|\Phi|^2\left(1-\frac{2|\Phi|^2}{\sigma^2}\right)^2
\end{aligned}
\label{eq:action}
\end{align}
where $R$ is the Ricci scalar. Our model ergostar is a BS composed of a complex scalar field $\Phi$ with mass $\mu$ and self-interaction strength $\sigma$, while $A_\mu$ is a complex vector field  with field strength $F^{\alpha\beta}$ that acts as the probe field triggering the instability. BSs are asymptotically flat, stationary, axisymmetric spacetimes with accompanying scalar field profile $\Phi\sim e^{-i(\tilde{\omega} t-\tilde{m}\varphi)}$ \cite{Kaup:1968zz, Ruffini:1969qy}; $\tilde{\omega}$ is the star's internal frequency, and $\tilde{m}$ its azimuthal index. We focus here entirely on the scalar theory with $\sigma=0.2$ and family of spinning BSs with $\tilde{m}=3$. Surfaces of constant scalar field magnitude $|\Phi|$ are tori. Star solutions are constructed numerically using the methods introduced in Refs.~\cite{Siemonsen:2020hcg,Kleihaus:2005me}. We consider the fastest growing ergoregion unstable massless vector field configurations; in the form $A_\mu \sim e^{-i(\omega t-m\varphi)}$, these are the $m=1$ states. We construct these linearly unstable states by solving the (complex) Maxwell equations starting from suitably chosen initial data on the fixed BS background (see Appendix~\ref{app:num_meth}). The frequencies and growth rates, $\omega=\omega_R+i\omega_I$, of these unstable states for four BSs along the sequence of solutions are shown in Fig.~\ref{fig:vector_freq}. We study the nonlinear saturation of the instability in the BS with, in units of total mass $M_0$, $\tilde{\omega}/\mu=0.427$, scalar mass $\mu M_0=3.85$, angular momentum $J_0/M_0^2=1.013$, and compactness $C=M_0/R_0=0.46$ \red{(where $R_0$ is the radius of the equilibrium star)}. This star has a pair of (un)stable polar light rings, as well as a pair of negative angular momentum equatorial ones (see the SM).

We numerically solve the complete set of Einstein-Klein-Gordon-Maxwell equations descending from \eqref{eq:action} in the time-domain. This is achieved by imposing axisymmetry on the metric, and enforcing $\mathcal{L}_\varphi\Phi=3i\Phi$ and $\mathcal{L}_\varphi A_\mu=iA_\mu$ (where $\mathcal{L}_\varphi$ is the Lie derivative in the direction of the azimuthal Killing vector $\varphi^\mu$) throughout the evolution using a generalized Cartoon method~\cite{Pretorius:2004jg}. This suppresses other non-axisymmetric instabilities known to exist in these star solutions~\cite{Sanchis-Gual:2019ljs,Siemonsen:2020hcg}, and allows us to study the ergoregion instability of $A_\mu$ in isolation. \red{Relaxing these symmetry assumptions would significantly increase the computational cost in our current setup (see Ref.~\cite{Siemonsen:2025fne} for a toy-model of the non-axisymmetric nonlinear evolution).} This also implies that the angular momenta of $\Phi$ and $A_\mu$ are \textit{separately} conserved, since $\nabla_\mu T^{\mu\nu}_A\varphi_\nu=0=\nabla_\mu T^{\mu\nu}_\Phi\varphi_\nu$, where $T_A^{\mu\nu}$ and $T_\Phi^{\mu\nu}$ are the energy-momentum tensors of the vector and scalar fields, respectively. Throughout, $J_A$ is the probe field's total angular momentum, obtained by integrating the density $\rho_J=n_\mu T^{\mu\nu}_A \varphi_\nu$,  where $n^\mu$ is the Cauchy surface normal, over a coordinate sphere of radius $20M_0$. In the above convention, this angular momentum is negative, $J_A<0$, for the linearly unstable field configuration (which is compensated by radiation carrying away positive angular momentum). The Einstein equations are solved in the Z4 formulation in moving puncture gauge~\cite{Bona:2003fj,vanMeter:2006vi}, while we assume Lorenz gauge for the vector field~\cite{Zilhao:2015tya}. For later convenience, we define the (Cauchy-slice dependent) projection $\chi=-n_\mu A^\mu$. The initial data consists of the stationary scalar BS described above, together with the most unstable massless vector field configuration. The addition of the vector field violates the Einstein constraints; however, we can control the effect of this, and the initial impact of other nonlinear effects, by choosing a sufficiently small initial amplitude.
To this end, we initialize the vector field with angular momenta $J^0_A$, at three different amplitudes corresponding to $-J^0_A/J_0\in\{0.004,0.007,0.012\}$. Further details regarding the numerical methods, with numerical convergence results, can be found in Appendix~\ref{app:conv}.

\section{Results}

\begin{figure}[t]
\includegraphics[width=1\linewidth]{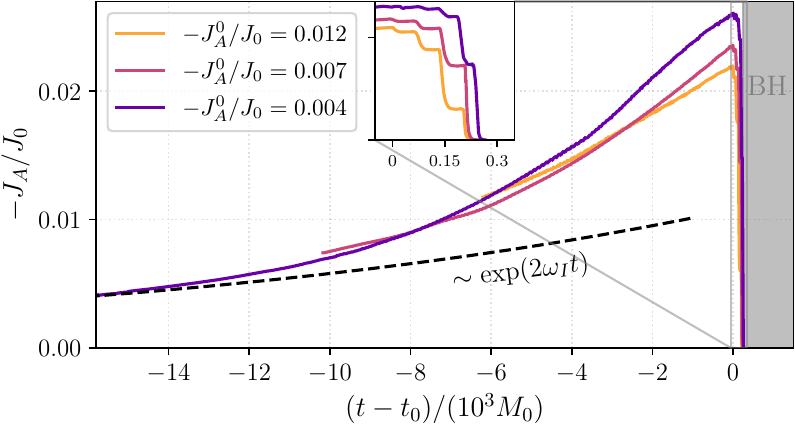}
\caption{Evolution of the probe field's angular momentum $J_A$, for three initial amplitudes. The curves are aligned in time at $t_0$, defined (roughly) as the peak of $|J_A|$. The dashed line indicates test field growth starting from $-J^0_A/J_0=0.004$.}
\label{fig:angular_momenta}
\end{figure}

\subsection{Black hole formation}

The final fate of the ergoregion instability in this system is black hole formation. We begin our discussion of the dynamics leading up to this complete collapse in the largely linear regime. In this phase, the system's evolution is dictated by the ergoregion unstable massless vector field growing exponentially, as can be seen in Fig.~\ref{fig:angular_momenta}. As the amplitude grows, we observe an enhancement of the instability's growth rate, denoted $\omega_I^{\rm NL}$, and real frequency, compared to the test field evolution. The former can be seen in Fig.~\ref{fig:angular_momenta} and follows roughly $\omega^{\rm NL}_I\approx \omega_I(1-114 J_A/J_0)$ (see also Appendix~\ref{app:conv}). This enhancement is likely driven by a weakly nonlinear shift of the linear frequency $\omega$ due to the backreaction of the massless vector $A_\mu$ on the geometry. This agrees with similar findings for Proca stars in Ref.~\cite{Siemonsen:2025wib}.

Eventually, the exponential growth phase ends and strongly nonlinear effects take over. This occurs in Fig.~\ref{fig:angular_momenta} around $t\approx t_0$. The system begins to shed the ergoregion unstable field, indicated by a rapid reduction of $|J_A|$ over a timescale of $\lesssim 200 M_0\ll 1/\omega_I$. At this time, an oscillation in the star with period $\approx 80 M_0$ is excited, leading to stages of vector field ejection, as can be seen in the inset of Fig.~\ref{fig:angular_momenta}. Recall, the unstable massless vector field mode is confined to reside predominantly around the negative angular-momentum equatorial stable light ring (see, e.g., Fig.~2 of Ref.~\cite{Siemonsen:2025wib}). The repeated ejection of this massless field signals a periodic breakdown and recovery of null trapping; that is, either the system efficiently injects energy to elevate the quasi-bound states into mostly radiating field modes, or the (effective) potential barrier trapping the massless field is repeatedly reduced. 

\begin{figure}[t]
\includegraphics[width=1\linewidth]{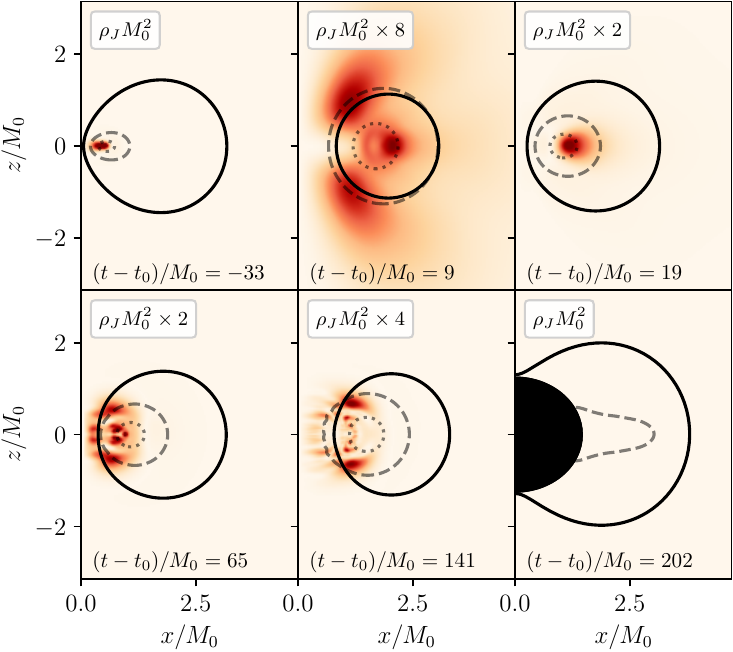}
\caption{The unstable field's angular momentum density $\rho_J$ at a few select times during the strongly nonlinear phase around $t_0$. The solid black lines indicate where $g_{tt}=0$ (the ergosurface, when the system is stationary with respect to $t$). Gray lines show surfaces with $|\Phi|/\max|\Phi(t=0)|=10^{-1}$ (dotted) and $10^{-3}$ (dashed).
For visual clarity, we rescale the angular momentum density in a few panels (as indicated by the labels). The apparent horizon's interior is approximately indicated by the black area. The $z$-axis is the symmetry axis.}
\label{fig:snapshots}
\end{figure}

To gain further understanding, in Fig.~\ref{fig:snapshots} we present a few snapshots of the evolution of the system during this highly nonlinear phase. The top left panel shows the state of the system beforehand, roughly during the exponential growth phase. There, the unstable field is confined to the ergoregion generated by the torus-shaped spinning BS, and is localized around the stable light ring. Strongly nonlinear effects begin to dominate around $t\approx t_0$ with large radial oscillations of the star. The top row of Fig.~\ref{fig:snapshots} shows the star oscillating out and away from the symmetry axis.\footnote{Note, gauge artifacts can cause a similar oscillatory motion~\cite{Evstafyeva:2025mvx}.} During each such oscillation, some of the ergoregion unstable field becomes untrapped and radiates towards null infinity, leading to a step-like reduction in the magnitude of the angular momentum $|J_A|$ inside the star. The first two panels in the bottom row show the state of the system just before launching the last two bursts of emission prior to black hole formation. At this stage, small spatial features develop in $\rho_J$, which are indicative of a turbulent direct cascade as discussed in more detail below. Finally, a black hole forms with mass $M_{\rm BH}/M_0\approx 0.94$ and dimensionless spin $J_{\rm BH}/M_{\rm BH}^2\approx 0.95$ (with numerical uncertainties of $\lesssim 10\%$). Recall, the probe field $A_\mu$ and the scalar matter $\Phi$ are unable to exchange angular momentum in our setup. As no significant vector field is absorbed by the black hole, angular momentum conservation dictates that some of the scalar matter is ejected prior to black hole formation, carrying away the remaining angular momentum. 

\begin{figure*}
    \centering
    \includegraphics[width=0.67\linewidth]{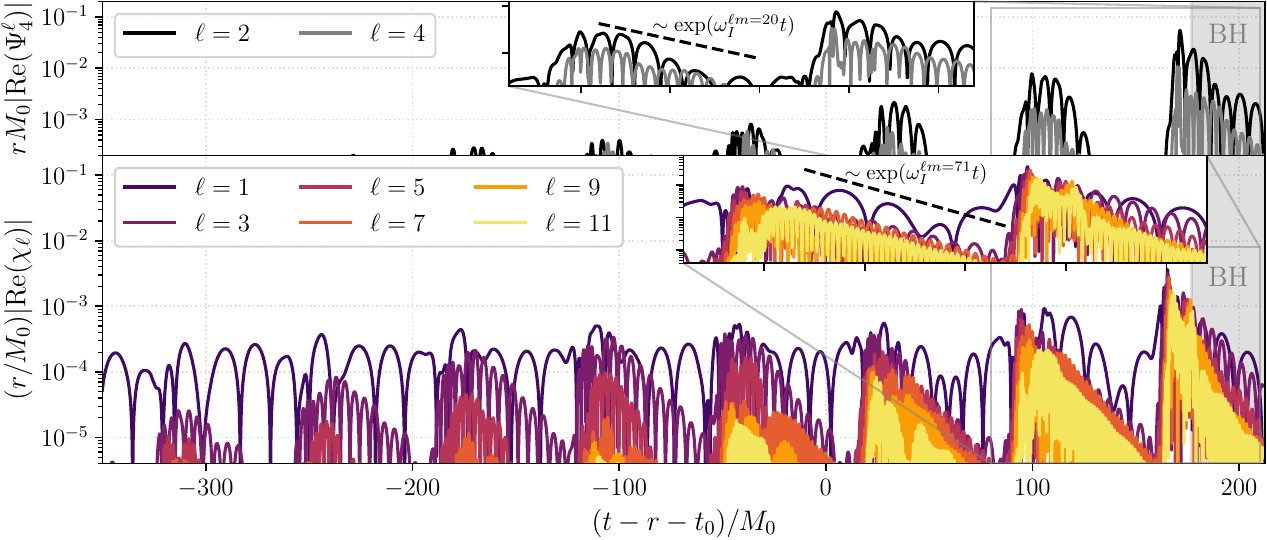}\hfill
    \includegraphics[width=0.30\linewidth]{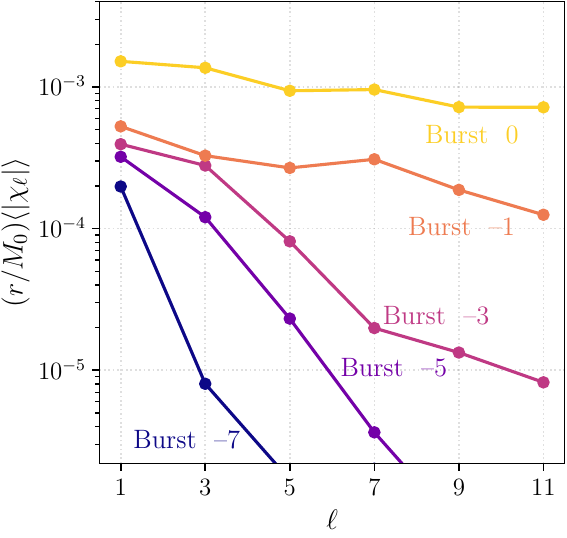}
    \caption{(left) The gravitational wave emission (top panel) and massless vector emission (bottom panel) for several different polar modes $\ell$ of the (spin-weighted) spherical harmonic decomposition of the Newman-Penrose scalar $\Psi_4$ and massless vector projection $\chi$ on a coordinate sphere of radius $r$, respectively. Recall, $\Psi_4$ contains only $m=0$ modes and $\chi$ only $m=1$ modes. In both insets, we also compare the burst decay timescales with the gravitational and electromagnetic quasi-normal modes of a Kerr black hole with spin $a/M=0.95$ \cite{Berti:2005ys,Berti:2009kk}. The frequencies roughly match those shown here for all $\ell$. The time $t-t_0$ an apparent horizon is found is indicated in gray (and labeled ``BH"). (right) The spectrum of (time-averaged) massless vector modes $\chi_\ell$ during the peak of burst, where ``Burst 0'' labels the last such burst in the sequence prior to black hole formation.}
    \label{fig:emissions}
\end{figure*}

\subsection{Gravitational wave emission}

In Fig.~\ref{fig:emissions}, we present the gravitational and massless vector wave emission around the strongly nonlinear phase of the instability's saturation before black hole formation. There are several features to note. First, in the weakly nonlinear regime (for $t\lesssim t_0$), we find the gravitational wave amplitude to be highly suppressed. At the same time, since the ergoregion instability of $A_\mu$ is driven by positive energy flux through the dominant $\ell=m=1$ emission channel, this dominates the massless vector energy flux during this phase of the system's evolution. Second, the star's radial oscillations trigger bursts of both gravitational wave and massless vector emission.\footnote{Bursts of massive scalar field are also emitted, but with subdominant energy fluxes.} In the strongly nonlinear regime (for $t\gtrsim t_0$) the amplitude of these bursts grows rapidly and begins dominating the emitted massless vector power. Thirdly, \textit{both} the frequencies and decay rates of the massless vector burst are comparable to the electromagnetic quasi-normal mode frequencies and decay rates of a Kerr black hole with spin $a/M=0.95$ for all polar modes (we have explicitly checked up to $\ell=7$ using results from Refs.~\cite{Berti:2005ys,Berti:2009kk}); see the inset of Fig.~\ref{fig:emissions}. While the gravitational wave frequencies of the $\ell=2$ and $\ell=4$ modes approximately match with the quasi-normal mode frequencies of such a spinning black hole, the decay of each burst only approximately resembles an exponential decay, at best (as can be seen in the inset of Fig.~\ref{fig:emissions}). The close resemblance of the emission to black hole quasi-normal modes can be understood qualitatively as follows: the release of the massless vector field trapped in the stable light ring occurs across the unstable light ring. As it travels through, the unstable light ring's characteristic frequencies imprint themselves on $\chi_\ell$. Since the spinning BS is highly compact, those frequencies and decay rates are close to those of a black hole (as shown explicitly in Ref.~\cite{Siemonsen:2024snb}). The total energy radiated in gravitational waves is $4\times 10^{-4} M_0$, while the massless vector field emits $\approx 0.02 M_0$ in the last $200M_0$ prior to black hole formation. Integrated throughout the evolution, the probe field radiates \textit{no net} angular momentum (as can be deduced from Fig.~\ref{fig:angular_momenta}).

\subsection{Gravitational turbulence}

Another feature encoded in the emission shown in Fig.~\ref{fig:emissions} is the appearance of a turbulent process transferring energy from $\ell=m=1$ towards high-$\ell$ vector states through a direct cascade active during the strongly nonlinear phase of the evolution prior to black hole formation. At a qualitative level, the amplification of higher-order polar modes can be seen in the angular momentum density shown in Fig.~\ref{fig:snapshots} (bottom row). These higher order polar states begin radiating in burst-like fashion (see the bottom left panel of Fig.~\ref{fig:emissions}) and their amplitude grows towards late times; in particular, high-$\ell$ modes grow more rapidly than lower-$\ell$ states. This implies that a spectrum of higher-order polar modes is built up during the strongly nonlinear evolution. This is explicitly shown in the right panel of Fig.~\ref{fig:emissions}. While early on, the emissions are dominated by the $\ell=m=1$ mode, with each burst the spectrum becomes supported over higher and higher polar modes. This ultimately causes the last burst to exhibit a practically flat spectrum $|\chi_\ell|$ (from $\ell=1$ up to $\ell=11$). Note, the gravitational wave emission shows no such transfer to higher-order polar modes. As $A_\mu$ lacks explicit nonlinearities, this turbulent process must necessarily be \red{mediated} by nonlinear \textit{gravitational} interactions.

\section{Discussion and Conclusion}

In this work, we numerically solved the full Einstein-Maxwell-Klein-Gordon system governing the evolution of the ergoregion instability of a spinning BS triggered by a massless vector field from the weakly nonlinear phase through saturation in axisymmetry. We find four key results: (i) a weakly nonlinear enhancement of the instability; (ii) radiation is emitted in distinctive bursts with increasing amplitude and with frequencies and decay rates comparable to those of a black hole with corresponding spin; (iii) nonlinear gravitational interactions transfer energy from the unstable mode to higher-order polar modes in a direct cascade; and (iv) the eventual fate of the star is to form a rapidly spinning black hole.

The ergoregion instability operates by having a growing mode in the ergoregion, compensated in energy and angular momentum---at the test field level---just by the radiation of the same field. In our case, the unstable massless vector field interacts with the star only through gravity, does not exchange angular momentum with the star, and eventually (in the nonlinear phase) essentially completely radiates away. Thus, in the weakly nonlinear regime, with approximately conserved scalar U(1)-charge, the unstable mode likely primarily taps into the gravitational binding energy, making the star more compact and increasing the instability rate and frequency; \red{this higher compactness lacks sufficient angular momentum support,} ultimately causing the collapse to a black hole. \red{Note, the turbulent cascade likely plays only a sub-dominant role in the black hole formation.} One can contrast this behavior with the backreaction of the superradiant instability of a massive scalar or vector field around a spinning black hole~\cite{East:2017ovw}. In that case, the horizon provides a way to dissipate the negative angular momentum (along with the negative energy) in the unstable field, and thus to spin down the black hole as the instability grows. In the present case of a horizonless ergostar, there is no such mechanism, and the star does not adiabatically spin down.

Given the results here, to what degree do they generalize to other systems? \red{One may \textit{speculate} that} since the only interaction between the unstable mode and the star is through gravity, one could expect ergostars composed of other matter to behave similarly \red{(e.g., in Ref.~\cite{Siemonsen:2025wib} an analogous enhancement of the instability in the weakly nonlinear regime was observed for Proca stars, signaling that strongly nonlinear effects dictate the final state there too).}
Likely more important would be what happens when one goes beyond axisymmetry. In that case, one would expect the fastest growing ergoregion instability to be associated with gravitational waves (analogous to $w$-modes in neutron stars~\cite{Kokkotas:2002sf}). 
Though not explicitly forbidden from exchanging angular momentum with the ergostar, one may still expect these trapped spacetime modes to be very weakly coupled to the matter~\cite{Kokkotas:2002sf}. Given this, or an approximately conserved bare mass of the matter making up the ergostar (originating from, e.g., a conserved U(1)-charge or baryon number), it seems plausible that in this case too, the ergoregion instability would take away gravitational binding energy, driving the star to be more compact. Since these objects are by construction ultra compact to begin with, it also seems plausible that the generic outcome of this is the collapse to a black hole. \red{A more systematic study of other objects is left to future work.} In the absence of a non-classical mechanism preventing horizon formation, this has strong implications for the existence of ultra compact horizonless objects.

The appearance of a direct turbulent cascade occurs when strongly nonlinear gravitational effects begin to be important. \red{To the best of our knowledge, this the first example of a direct cascade of this kind in asymptotically flat spacetimes (for details in asymptotically Anti-de Sitter spacetimes see, e.g., Refs.~\cite{Bizon:2011gg,Buchel:2012uh,Dias:2011ss,Niehoff:2015oga}).} This cascade is, however, largely restricted to the perturbing massless vector field. In particular, \red{we find no strong evidence for} small-scale features in curvature invariants, since even at its peak, the perturbing field only contains $\sim 1\%$ of the star's energy. As a result of this, during the nonlinear phase prior to black hole formation, the turbulent cascade strongly impacts the evolution of the massless unstable field only.
It is, however, largely irrelevant for the evolution of the spacetime; that is, black hole formation is induced likely by large-scale and low-frequency perturbations. Therefore, gravitational nonlinearities induce weak vector self-interactions, which then lead to the amplification of higher-order polar modes. This is qualitatively consistent with the results of Ref.~\cite{Siemonsen:2025fne}, where the appearance of turbulence had been observed in a scalar toy model for the Einstein equations. Similar to speculations there, the turbulent process here may also be triggered by parametric resonances, which may explain the short amplification timescale $\sim\mathcal{O}(10^2)M_0\ll 1/\omega_I$. 

As argued in Ref.~\cite{Siemonsen:2025fne}, outside of axisymmetry, the gravitational wave driven ergoregion instability should cause the population of higher-order azimuthal modes (in contrast to polar vector states as presented here) through a direct turbulent cascade. Unless the resulting turbulent spectrum exhibits stronger support at small scales compared to large ones (which is observed neither here, nor in Ref.~\cite{Siemonsen:2025fne}), the spacetime evolution is dominated by large-scale dynamics, causing black hole formation as discussed above. 

Finally, the burst-like character of the radiation found here is reminiscent of so-called gravitational wave echoes~\cite{Cardoso:2016oxy,Abedi:2020ujo}. Particularly, the close alignment of both frequencies and decay rates of the emitted signal and quasi-normal mode frequencies of a black hole with the corresponding spin is striking. However, unlike echoes---which are purely a propagation effect of test fields in the stable light ring---the bursts likely arise here as large-scale oscillations of the bulk of the spacetime allow the unstable mode that was trapped at the test field level to become partially untrapped.
In the non-axisymmetric case, where the ergoregion instability grows through gravitational radiation, one expects a quasi-monochromatic, exponentially growing gravitational wave signal (in the linear regime), to eventually give way to a burst-like signal with imprints of the nonlinear dynamics and gravitational turbulence. 
Naturally, this may serve as a smoking-gun signature of black hole mimickers and has implications for gravitational wave searches for ultra compact objects.

\section{Acknowledgements}

We thank Tamara Evstafyeva and Frans Pretorius for fruitful discussions.
W.E. acknowledges support from a Natural Sciences and Engineering Research
Council of Canada Discovery Grant and
an Ontario Ministry of Colleges and Universities Early Researcher Award.
This research was 
supported in part by Perimeter Institute for Theoretical Physics. Research at
Perimeter Institute is supported in part by the Government of Canada through
the Department of Innovation, Science and Economic Development and by the
Province of Ontario through the Ministry of Colleges and Universities.
This work used \texttt{anvil} at Purdue University through allocation PHY250024 from the Advanced Cyberinfrastructure Coordination Ecosystem: Services \& Support (ACCESS) program \cite{access}, which is supported by National Science Foundation (NSF) Grants OAC-2138259, -2138286, -2138307, -2137603, and -2138296. This research was enabled in part by support provided by the Digital Research Alliance of Canada (alliancecan.ca). Calculations were performed on the Symmetry cluster at Perimeter Institute and the Narval cluster at the École de technologie supérieure in Montreal.

\appendix

\section{Light ring structure} \label{app:light_rings}

\begin{figure}
    \centering
    \includegraphics[width=1\linewidth]{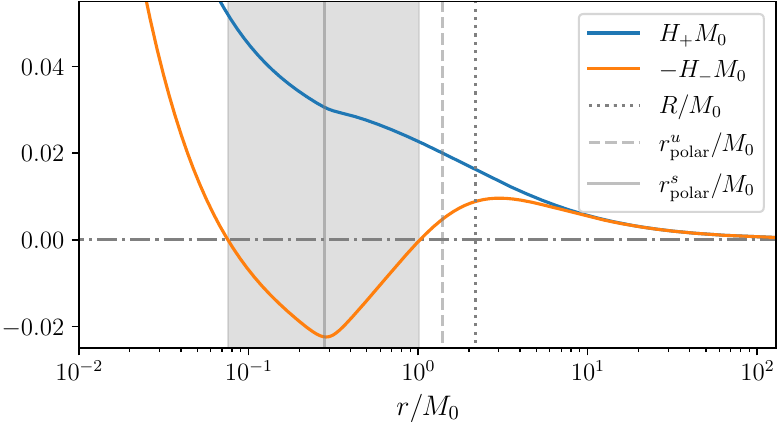}
    \caption{The light ring structure of the BS solution considered throughout the main text: $R$ is the star's radius (as defined in Ref.~\cite{Siemonsen:2020hcg}), $r^u_{\rm polar}$ ($r^s_{\rm polar}$) is the radial coordinate at which unstable (stable) polar light rings intersect the equatorial plane, $H_\pm$ are the effective potentials for equatorial null geodesics (defined in the text), and the gray shaded area is the coordinate size of the ergoregion in the equatorial plane.}
    \label{fig:light_rings}
\end{figure}

We briefly investigate the structure of the light rings in the specific BS considered throughout this work. To that end, we utilize the ray-tracing methods from Ref.~\cite{Siemonsen:2024snb} to determine the locations of the stable and unstable polar light rings (i.e., trapped null geodesics with vanishing angular momentum with respect to the axis of symmetry). Equatorial light rings are identified by means of the effective potentials $H_\pm$ for positive and negative angular momentum null geodesics, respectively~\cite{Cunha:2017qtt} (see also Ref.~\cite{Cunha:2016bjh}).
In Fig.~\ref{fig:light_rings}, we show both of these and compare to the star's surface (defined as $R=M_0/C$) and location of the ergoregion. From this, we can explicitly see that there are a pair of polar light rings---one stable and one unstable, as well as a pair of equatorial light rings for negative angular momentum null geodesics (implied by the existence of an ergoregion). The spacetime is, however, not sufficiently compact to exhibit also positive angular momentum equatorial light rings, since $H_+$ lacks local extrema.

\section{Numerical methods} \label{app:num_meth}

\subsection{Evolution}

The stationary spinning BSs considered in this work are obtained numerically using the techniques outlined in Refs.~\cite{Siemonsen:2020hcg,Kleihaus:2005me}. The Einstein equations are numerically evolved using the (3+1)-decomposition of the Z4 formulation~\cite{Bona:2003fj} in the ``shifting-shift" version of the moving puncture gauge~\cite{vanMeter:2006vi} (with damping $\eta=4/M_0$), while imposing axisymmetry on the metric using a Cartoon method~\cite{Pretorius:2004jg}. 

The complex vector field equations are solved numerically in the time-domain using the formulation introduced in Refs.~\cite{Zilhao:2015tya,East:2017ovw}, assuming the Lorentz gauge $\nabla_\mu A^\mu=0$, and imposing the $m=1$ azimuthal symmetry $\mathcal{L}_\varphi A_\mu=iA_\mu$. Similarly, the scalar field is evolved numerically using techniques developed in Ref.~\cite{Siemonsen:2020hcg}, while imposing $\mathcal{L}_\varphi \Phi=3i\Phi$. We use constraint damping with parameters $\kappa=2/M_0$ and $\rho=-0.85$ (as defined in Ref.~\cite{Gundlach:2005eh}). All equations are discretized using fourth order accurate finite difference stencils and employing a fourth order accurate Runge-Kutta time stepping. Throughout most of the evolution, the star is covered with 12 levels of adaptive mesh refinement with refinement ratio 2:1 and grid spacing of $h=8\times 10^{-4}M_0$ on the finest level. During the last few $\mathcal{O}(10^2) M_0$ prior to black hole formation, the adaptive mesh refinement is allowed to add up to 2 additional levels to resolve the turbulent process and small spatial features. Using the Z4 formulation without any conformal decomposition of the metric,
we are not able to evolve for very long times after the formation of a black hole, likely due to large unresolved gradients near the region of collapsing lapse (the lapse is floored at $\alpha_0=10^{-4}$); depending on resolution, the evolution fails between $25M_0$ and $50M_0$ after confidently finding the apparent horizon. Due to this, we extract the gravitational and vector radiation at relatively small radii as detailed below. Lastly, in the
case that we evolve the longest ($-J^0_A/J_0=0.004$), we find that over long timescales the lapse begins to decrease substantially away from unity at large distances from the star. While we have not identified the origin of this drift, this gauge evolution may be caused by accumulation of (largely unresolved) radiation in the regions close to spatial infinity on our compactified numerical grid. To mitigate the impact on diagnostics, we restart this evolution at $t-t_0=-3.7\times 10^3 M_0$ and use a radial transition functions to smoothly set the lapse to unity roughly outside radial coordinate radius of $200 M_0$. This causes a small---and for our results insignificant---change in the growth rate of $|J_A|$ presented in the main text.

\subsection{Initial data}

\begin{figure}[t]
\includegraphics[width=0.48\textwidth]{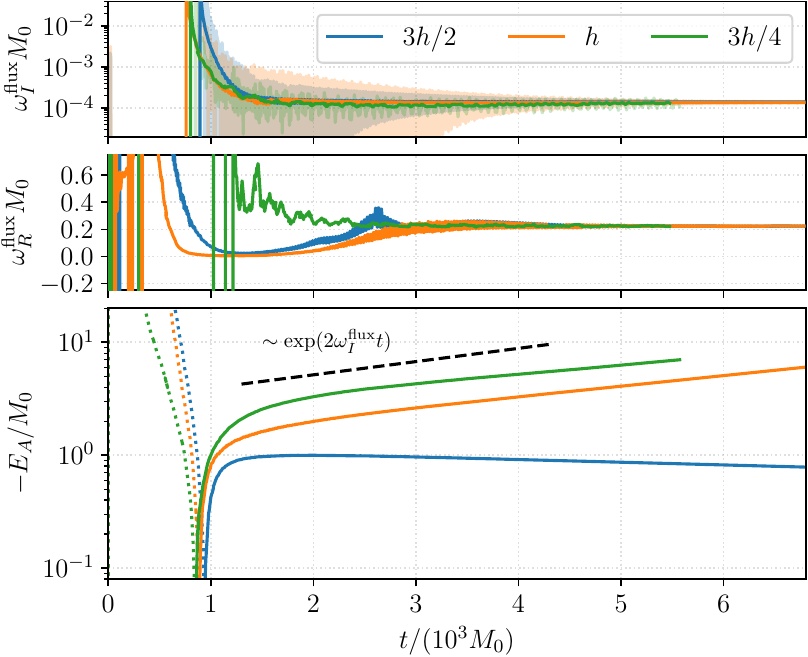}
\caption{The evolution of $\omega_R^{\rm flux}$ and $\omega_I^{\rm flux}$, as well as the total energy $E_A$ (with arbitrary normalization), starting from the noise initial data described in the text for different resolutions compared to the default grid spacing $h$. In the top panel, we show also the ratio $P M_0/L$ (no time average applied) as shaded lines. In the bottom panel, dotted lines correspond to positive $E_A$.}
\label{fig:vector_conv}
\end{figure}

To generate suitable initial data for the massless vector field we proceed as follows: on a fixed BS background all vector components, $A_\mu=\chi_\mu+n_\mu \chi$, are set to zero outside the ergoregion and set to a point-wise random number in the interval $[-1,1]$, where $g_{tt}>0$. 
We evolve the vector test-field equations of motion for
these data for $\sim\mathcal{O}(10^3)M_0$ until an exponentially growing mode appears. This mode is assumed to be the fastest growing ergoregion unstable field configuration (recall, $m=1$ is enforced explicitly). To that end, we define the vector field's energy, $E_A$, by the integral of $n_\mu T_A^{\mu\nu}t_\nu$ out to coordinate radius $20M_0$ (here $t^\mu$ is the asymptotically timelike Killing field), as well as the associated energy and angular momentum fluxes through this coordinate sphere by $P$ and $L$, respectively. As an example, in Fig.~\ref{fig:vector_conv} we present the evolution of the vector field's total energy $E_A$, as well as measures of the real $\omega^{\rm flux}_R=\langle P\rangle M_0/\langle L\rangle$, and imaginary $\omega^{\rm flux}_I=-M_0\langle P\rangle/(2E_A)$, parts of this linear ergoregion unstable field configuration. 
Here $\langle\dots\rangle$ denotes a moving time average. After an initial transient, the field configuration settles into an exponentially growing state with constant frequency and growth rate. The values quoted in the main text are then simply $\omega_R=\omega_R^{\rm flux}$ and $\omega_I=\omega_I^{\rm flux}$. The error bars presented in the main text are obtained as in Ref.~\cite{Siemonsen:2025wib}: the difference between $\omega_I^{\rm flux}$ and an exponential fit to $E_A$ measures the numerical uncertainties in the growth rate, and the amplitude of the oscillations in $P M_0/L$ compared to $\omega_R^{\rm flux}$ quantifies the uncertainty of our method to determine the real frequency. Together with the stationary BS solutions, the final state of this evolution, at around $t=6\times 10^3 M_0$, [together with explicitly enforcing the equatorial symmetry $E^z(-z)=-E^z(z)$, $\chi^z(-z)=-\chi^z(z)$, as well as $E^i(-z)=E^i(z)$ and $\chi^i(-z)=\chi^i(z)$ for $i\in\{x,y\}$ and $\chi(z)=\chi(-z)$] is used as initial data for the fully nonlinear evolution of the Einstein-Klein-Gordon-Maxwell equations.

\section{Convergence and accuracy}
\label{app:conv}

\subsection{Constraint violations}

\begin{figure}
    \centering
    \includegraphics[width=1\linewidth]{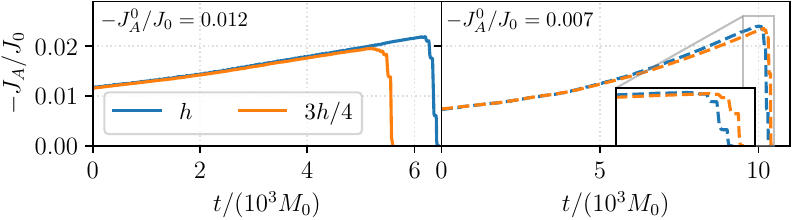}
    \includegraphics[width=1\linewidth]{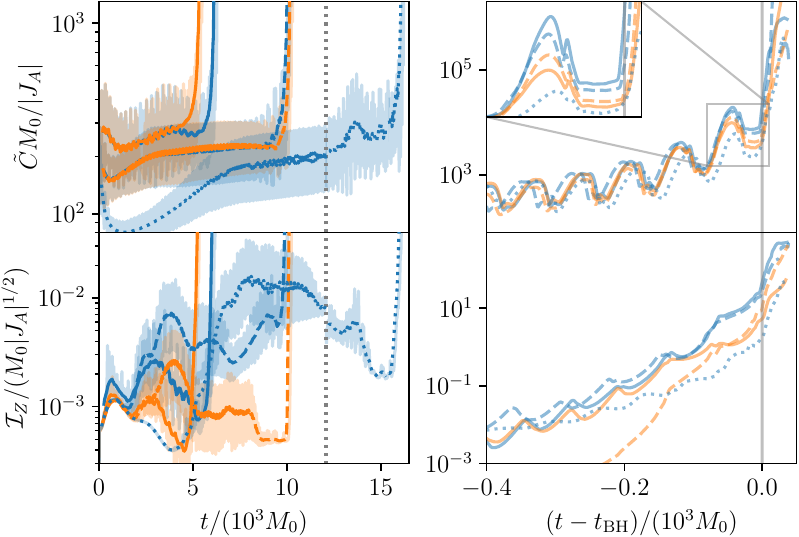}
    \caption{ Top: The evolution of $J_A$ starting from two initial amplitudes at two different resolutions with respect to the default grid spacing $h$. The evolutions are aligned at $t=0$. Bottom left column: The integrated Einstein constraint violations (top) and vector constraint violations (bottom) normalized by the unstable mode's angular momentum throughout the exponential growth phase (shaded curve). Unshaded lines are moving time-averages. The colors correspond to the different resolutions, while different initial amplitudes are indicated with different line styles: $-J^0_A/J_0=0.012$ (solid), $-J^0_A/J_0=0.007$ (dashed), and $-J^0_A/J_0=0.004$ (dotted). All curves are aligned at $t=0$, and the gauge switch time of the $-J_A^0/J_0=0.004$ case is indicated as a vertical gray dotted line. Bottom right column: Same as the left column, but with curves aligned at the time the apparent horizon is first found $t_{\rm BH}$.}
    \label{fig:constraints}
\end{figure}

To estimate the accuracy of our methods, and establish the validity of our results, we perform higher-resolution evolutions with grid spacing $3h/4$ for the two cases with largest initial amplitude, i.e., $-J^0_A/J_0\in\{0.012,0.007\}$ (in addition to the default resolution of $h=8\times10^{-4}M_0$). We begin by quantifying the violation of the Einstein constraints, as well as monitoring the $Z$-field damping violation of the Gauss's Law constraint $D_i(\gamma^i_\mu F^{\mu\nu}n_\nu)=0$ according to Ref.~\cite{Zilhao:2015tya} (here $\gamma_{\mu\nu}=g_{\mu\nu}+n_\mu n_\nu$ and $D_i$ is the covariant derivative with respect to $\gamma_{ij}$). In particular, we define the integrated Einstein constraint violation by
\begin{align}
    \tilde{C}=\int_Bd^3x\sqrt{\gamma}\left(\mathcal{H}^2+\sum_{i=1}^3\mathcal{M}_i^2\right)^{1/2},
\end{align}
where $\mathcal{H}$ and $\mathcal{M}_i$ are the Hamiltonian and momentum constraints, respectively, and $B$ is a coordinate ball of radius $20M_0$. Similarly, we define $\mathcal{I}_Z=\int_B d^3x\sqrt{\gamma} |Z|$. While the background BS solution satisfies the Einstein constraint equations, the ergoregion unstable vector field configuration does not. These violations scale as $\mathcal{H},\mathcal{M}_i\sim T^A_{\mu\nu}\sim J_A$. Non-vanishing $\mathcal{I}_Z$ is due to truncation error of the evolution used to obtain these initial data. In Fig.~\ref{fig:constraints}, we present the behavior of these quantities throughout all evolutions performed in this work. During the exponential growth phase, the Einstein constraint violations are driven by the initial violation introduced by the vector field and scales with its square amplitude (as measured by $J_A$). We find that $\tilde{C}M_0/|J_A|$ decreases with decreasing initial amplitude $J^0_A$ by roughly a factor of two between $-J^0_A/J_0=0.012$ and $-J^0_A/J_0=0.004$. Recall, in the $-J_A/J_0=0.004$ case, we changed the gauge evolution at $t\approx 12\times 10^3 M_0$, leading to a slightly different subsequent evolution of the gauge-dependent measures $\mathcal{I}_Z$ and $\tilde{C}$. Just prior to, and after, black hole formation, $\tilde{C}$ appears to no longer be driven by the constraint violations from the initial data, but rather numerical truncation error from the evolution (see left column of Fig.~\ref{fig:constraints}). At $t=t_{\rm BH}$,  $\tilde{C}$ converges towards zero at roughly second order (a slight dephasing of the star's oscillations between resolutions renders a comparison across a larger time interval challenging). The vector constraint is better behaved throughout, as constraint damping results in a significant reduction of $\mathcal{I}_Z$ in the high- compared with low-resolutions evolutions. However, prior to black hole formation, this quantify is also rapidly amplified, as the system becomes turbulent and small-scale features are formed. Note, around black hole formation $J_A$ approaches zero (as the vector field is emitted to null infinity), leading to an artificial up-tick of the measures shown in the right panels of Fig.~\ref{fig:constraints} (in particular, $\tilde{C}/M_0$ decreases after $t_{\rm BH}$).

\subsection{Weakly nonlinear effects}

\begin{figure}[t]
    \centering
    \includegraphics[width=1\linewidth]{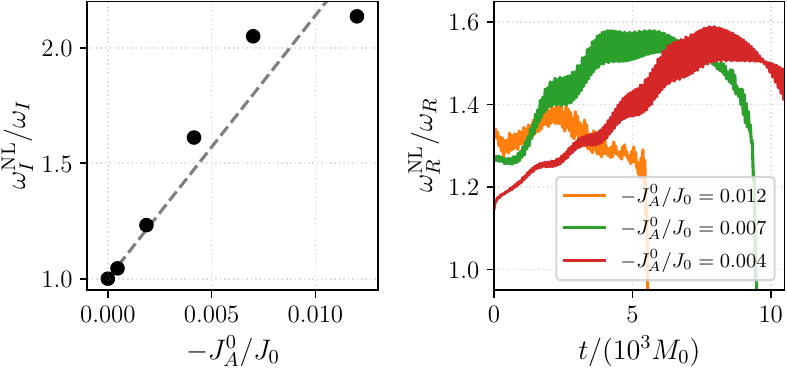}
    \caption{Left: The growth rate $\omega_I^{\rm NL}$ of the unstable field configuration (normalized by its test field rate $\omega_I$) as a function of its total angular momentum. For reference, the dashed line is the linear fit $-114J_A^0/J_0+1$ through the two data points with lowest non-zero $J_A^0$. This is analogous to Fig.~8 in Ref.~\cite{Siemonsen:2025wib}. Right: The measure $E_A/J_A$ of the real frequency of the $m=1$ field configuration, normalized by the test field frequency $\omega_R$, as a function of time for the cases evolved all the way through the nonlinear phase.}
    \label{fig:weakly_nl_freq}
\end{figure}

As noted above, we find a large nonlinear enhancement of both the growth rate and real frequency of the unstable massless vector field configuration compared with the linear growthrate $\omega_I$ and frequency $\omega_R$.
Here, we further quantify the amplitude dependence of this enhancement.
To that end, we consider evolutions of varying initial amplitude $-J^0_A/J_0\in\{4.7\times 10^{-4},1.9\times 10^{-3},0.004,0.007,0.012\}$ and compare different means of determining both the real and imaginary frequencies of the resulting vector field configurations. In particular, we compare $\tilde{\omega}^{\rm NL}_I=-L/(2J_A)$ with $\omega_I^{\rm NL}$ obtained from an exponential fit to $-J_A(t)$ (at early times), as well as $\tilde{\omega}_R^{\rm NL}=PM_0/L$ with $\omega_R^{\rm NL}=E_A/J_A$.

We find consistency among these methods of computing the weakly nonlinear corrections to the test field frequency and growth rate to within $\sim 10\%$ relative difference, across different resolutions, and find a temporal increase of $\omega_R^{\rm NL}$ and $\omega_I^{\rm NL}$ in the $-J^0_A/J_0\in \{0.007,0.004\}$ cases. We illustrate this in Fig.~\ref{fig:weakly_nl_freq}. Specifically, the left panel demonstrates explicitly that the evolutions with $-J^0_A/J_0\in \{0.007,0.004\}$ are in the weakly nonlinear regime since $\omega^{\rm NL}_I$ scales roughly linearly with $J_A^0$ as expected~\cite{Siemonsen:2025wib}. The right panel shows that $\omega_R^{\rm NL}$ increases throughout the evolution, as $-J_A(t)$ grows towards saturation. 

\subsection{Radiation}

\begin{figure}[t]
    \centering
    \includegraphics[width=0.48\linewidth]{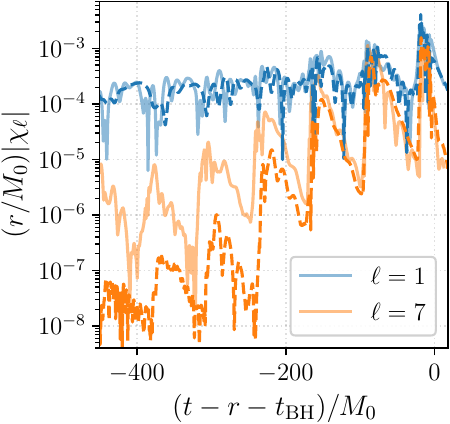}\hfill
    \includegraphics[width=0.49\linewidth]{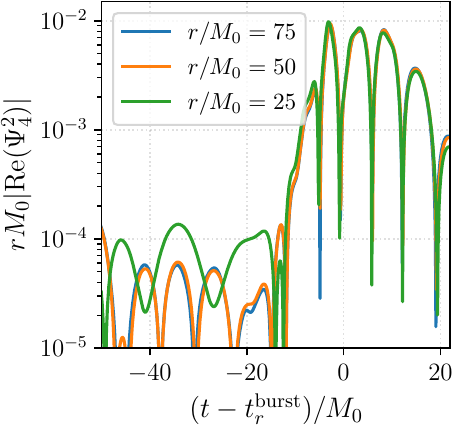}
    \caption{Left: A comparison of the evolution of the $m=1$ spherical harmonic of $\chi$ on a sphere of radius $20M_0$ between the default grid spacing $h$ (solid lines) and $3h/4$ (dashed lines) for the case with initial amplitude $-J^0_A/J_0=0.007$. Right: Comparing the GW emission extracted through a spin-weighted spherical harmonic decomposition of $\Psi_4$ on coordinate spheres of radii $r/M_0\in \{25,50,75\}$. We focus here on the second-to-last burst in the sequence. The waveforms are aligned in time at $t=t^{\rm burst}_r$.}
    \label{fig:radiation_conv}
\end{figure}

Finally, we analyze the accuracy of our predictions for the gravitational and massless vector radiation. Focusing first on the vector emission, in Fig.~\ref{fig:radiation_conv} we show the resolution dependence of the turbulent amplification of higher-order polar modes for the case with $-J^0_A/J_0=0.007$. As evident from there, the amplitude of the bursts, the turbulent amplification timescale, and also the frequency (not shown) all are consistent across resolutions (and also with the evolutions with $-J^0_A/J_0\in\{0.012,0.004\}$). Wavelengths associated with the $\ell=7$ mode are well-resolved (on the sphere with radius $20M_0$, where $\chi_\ell$ is defined), such that the resolution dependence of the $\ell=7$ mode before turbulent amplification can be explained by higher levels of truncation error in the lower resolution evolution. While not shown here, we find the gravitational wave burst emission to be similarly consistent across different resolutions and initial amplitudes; specifically, the burst separation time, the burst's primary frequency, and the overall envelope, are consistent, while smaller details may differ. In addition, here we also verify that the extraction of $\Psi_4$ on spheres of relatively small radius of $25M_0$ is sufficient. In the right panel of Fig.~\ref{fig:radiation_conv}, we show the second-to-last burst prior to black hole formation in the $-J^0_A/J_0=0.007$ case, extracted on sphere with three different radii. From there we conclude that the burst features dominating the gravitational waveform are largely not impacted by the relatively small extraction radius. 

\bibliography{bib.bib}

\end{document}